\documentclass[aps,prl,amsfonts,twoside,twocolumn,amssymb,superscriptaddress]{revtex4}
\pdfoutput=1
\usepackage[latin1]{inputenc}

\usepackage[T1]{fontenc}
\usepackage{amsmath,amstext,amsfonts,amsxtra}
\usepackage{amsthm}
\usepackage{graphicx}
\usepackage{mathrsfs}

\usepackage{bbm}

\newcommand\bra[1]{\ensuremath{\langle{#1}\vert}}
\newcommand\ket[1]{\ensuremath{\vert{#1}\rangle}}
\newcommand\bracket[2]{\ensuremath{%
    \langle{#1}\mkern1.2mu\vert\mkern1.2mu{#2}\rangle}}
\newcommand\bracketOp[3]{\ensuremath{%
\langle{#1}\mkern1.2mu\vert #2 \vert\mkern1.2mu{#3}\rangle}}

\newcommand\ketbra[2]{\ensuremath{%
    \vert{#1}\mkern1.2mu\rangle\langle\mkern1.2mu{#2}\vert}}

\newcommand\hilbertH{\ensuremath{\mathscr{H}}}

\newcommand\hilbertE{\ensuremath{\mathscr{E}}}
\newcommand\hilberteins{\ensuremath{\mathbbm{1}}}

\newcommand\reellR{\mathbbm{R}}

\newcommand\intMenge[2][1]{\{#1,\dots, #2\}}

\newcommand\hilbertBOp[1][\hilbertH]{\ensuremath{\mathscr{B}(#1)}}

\DeclareMathOperator{\spa}{span}

\DeclareMathOperator{\tr}{tr}

\newcommand{\Betrag}[1]{\vert #1\vert}

\newcommand{\Menge}[1]{\left\{#1\right\}}

\theoremstyle{plain}
\newtheorem{lem}{Lemma}

\begin{document}
\title{Quantum cryptography as a retrodiction problem}
\author{A.~H. Werner\footnote{albert.werner@itp.uni-hannover.de}} \address{Institut f\"{u}r Theoretische Physik, Leibniz Universit\"{a}t Hannover,
    	Appelstra{\ss}e 2, 30167 Hannover}
\author{T. Franz} \address{Institut f\"{u}r Theoretische Physik, Leibniz Universit\"{a}t Hannover,
    	Appelstra{\ss}e 2, 30167 Hannover}
\author{R.~F. Werner} \address{Institut f\"{u}r Theoretische Physik, Leibniz Universit\"{a}t Hannover,
    	Appelstra{\ss}e 2, 30167 Hannover}

\begin{abstract}
We propose a quantum key distribution protocol based on a quantum retrodiction protocol, known as the Mean King problem. The protocol uses a two way quantum channel. We show security against coherent attacks in a transmission error free scenario, even if Eve is allowed to attack both transmissions. This establishes a connection between retrodiction and key distribution.
\end{abstract}

\maketitle

\section{Introduction}
Quantum key distribution (qkd) protocols allow two parties, traditionally called Alice and Bob, to generate a secret key, which enables them to communicate secretly via onetime pad encryption. There are qkd protocols, that guarantee the security of the key against an eavesdropper, who is capable of implementing arbitrary quantum operations (coherent attacks)
 \cite{lit:bb84,lit:renner_gisin,lit:shorpresc}.

After the execution of a qkd protocol Alice and Bob should share an identical and random bit string, which is unknown to a potential eavesdropper Eve. If one forgets for a moment about the secrecy of the bit string, the goal in a qkd protocol is that one of the communicating partners infers a bit value corresponding to a measurement outcome obtained or a state prepared by the other.
In an optimal protocol one would request that this is possible in every run of the protocol.

A similar problem has been proposed by Vaidman, Aharonov and Albert in the context of the retrodiction of a measurement result of a spin $\frac{1}{2}$-particle, known as the Mean King problem \cite{lit:VaidAhranov}. Alice in this setting has to guess the outcome of a measurement performed by Bob without knowing the measurement bases used. There has been a first proposal to use this setup in a quantum cryptographic context in \cite{lit:Bub}, but a security proof accounting for an arbitrary attack on both quantum channels was not given. In this paper we show, that there are solutions to the Mean King problem, which guarantee the security of these measurement results against an eavesdropper in a stronger scenario, thus establishing the connection between retrodiction and security. In addition the proposed protocol generates a bit of raw key in every single run.

\section{Setting and Result}
\subsection{The Mean King retrodiction problem}%
\label{sec_retrodiction}
The retrodiction problem can be stated as a quantum game played by two players. One player, Alice, wins the game if she can guess a measurement outcome obtained by the other player, Bob.
Beforehand both of them agree on a Hilbert space $\hilbertH$ of dimension $d$ and  a set of $d+1$ orthonormal bases of this Hilbert space $\Menge{\Phi_b(i);\; i=1, \dots, d,\, b=1, \dots, d+1}$. Here $\Phi_b(i)$ denotes the i-th basis vector of the b-th basis.

Alice starts the game by preparing a maximally entangled state $\Omega\in\hilbertH\otimes\hilbertH$ and sends the second system to Bob. He performs a projective measurement in a randomly chosen basis $b\in\intMenge{d+1}$, but keeps this choice and his measurement result $i$ secret. After Bob has returned the resulting eigenstate of his measurement to Alice, she holds the state (conditional on Bob obtaining $i$)
\begin{equation}\label{Phihat}
  \hat\Phi_b(i)=(\hilberteins\otimes\ketbra{\Phi_b(i)}{\Phi_b(i)})\Omega.
\end{equation}
After Alice has performed a final measurement $\Menge{F_x}$ on this state, obtaining a classical result $x$, all quantum information is discarded. In the last step Bob reveals his choice of basis $b$ and depending on this value and her measurement result $x$, Alice has to guess Bob's measurement outcome $i$.
We note that the precise form of the result $x$ is not important and we might think of it as a $d+1$-tupel $x=(x(1),\dots,x({d+1}))$ indicating to Alice that she should guess $i=x(b)$ if the basis $b$ was chosen by Bob. We refer to $x$ as a guessing function. As there are $d$ different possible measurement outcomes for each of the $d+1$ measurement bases, the set $X$ of possible guessing functions has $d^{d+1}$ different elements.
A successful strategy for Alice consists of a maximally entangled state $\Omega$ and a measurement $\Menge{F_x}$ such that the probability for a wrong guess is zero, which means that $\tr({F_x\ketbra{\hat\Phi_b(i)}{\hat\Phi_b(i)}})=0$, unless $x(b)=i$ and that she can make a guess in every round, implying the normalization condition $\sum F_x = \hilberteins$.

The existence of such a strategy has been studied in the case of mutually unbiased bases (MUBs) \cite{lit:VaidAhranov,lit:MUBs}. In \cite{lit:MeanerKing} it has been shown, that for constructing a winning strategy weaker conditions are sufficient. A set of $k$ bases only has to be {\em non-degenerate} meaning that the span of the projectors $\ketbra{\Phi_b(i)}{\Phi_b(i)}$ is $k(d-1)+1$ dimensional and has to {\em admit a classical model}. Here a set of $k$ bases is said to admit a classical model, if there exists a probability distribution of $k$ variables, each taking $d$ values, such that its marginals equal the probability distributions of the joint probabilities $p_{ab}(i,j)=\frac{1}{d} \Betrag{\bracket{\Phi_b(i)}{\Phi_a(j)}}^2$ for all pairs of bases.

In our scenario with $k=d+1$, non-degeneracy implies also that the projectors span the space of all hermitian operators on $\hilbertH$.
Of course, $d+1$ mutual unbiased bases (MUBs) are an example of a set of bases exhibiting this properties \cite{lit:MeanerKing}.

\subsection{The qkd protocol}
\label{sec_protcol}
We assume a setting for the Mean King problem, in which the set of bases is non-degenerate and Alice has a successful strategy, that is maximal in the sense of incorporated measurement projectors (see bellow). In each run, there will be $n$ instances of the Mean King problem, and we use a vector arrow to indicate $n$-tuples of choices outcomes, etc. It is irrelevant for our analysis, whether the steps are carried out sequentially for each instance, or for the full block of $n$ instances simultaneously.

\begin{enumerate}
\item Initially, an entangled state $\rho_n$ of $n$ pairs is generated and distributed to Alice and Bob. This process is vulnerable to attack, but they proceed as if
    $\rho_n=\ketbra{\Omega^{\otimes n}}{\Omega^{\otimes n}}$ consists of $n$ copies of the pure state used for the Mean king problem.
\item Bob chooses at random $n$ measurement bases $\vec b$, performs the projective measurement with basis $b_k$ on the $k^{\rm th}$ particle, obtaining the results $i_k$, and sends back the corresponding eigenstate. Altogether he returns  $\Phi_{\vec i}(\vec b)=\bigotimes_{k=1}^n\Phi_{i_k}(b_k)$, without disclosing $\vec i$ or $\vec b$.
\item Alice performs her measurement $\{F_x\}$ on each of the returning particles and the corresponding one in her storage, obtaining as a result an $n$-tuple $\vec x$ of guessing functions.
\item After Alice announces that she finished the last of her measurements Bob publishes his choice of bases $\vec b$ from which Alice infers $i_k^\prime=x_k(b_k)$.
\end{enumerate}
Since this is a Mean King game between Alice and Bob, without Eve's interaction this will produce an identical string $\vec i'=\vec i$ of $d$-digits. In order to test for the presence of an eavesdropper, Alice and Bob will randomly select some particles $k$ and check for the agreement $i'_k=i_k$ in these instances, and accept, if they never find a deviation.

\subsection{Security}
The {\em transmission error free scenario} for the security analysis assumes that Alice and Bob do find agreement with probability 1, i.e., that a potential attacker does not risk the introduction of any errors at all, and also that there are no spontaneous transmission errors. Of course, a full analysis would have to allow for errors, so the proof of security in this scenario is only a proof of principle.

In our setting, Eve can interact at different stages. First she may provide the initial state, possibly keeping a system of her own entangled with the distributed pairs. Then she may interact with the states that Bob returns to Alice in a coherent way and finally making  a joint measurement on her system. Since we analyse whole blocks simultaneously, we even allow choices violating time ordering.

\emph{What we show in the next section is that if Eve's actions do not interfere with the perfect key agreement, which Alice and Bob can test in principle, then her final conclusion will be uncorrelated with the key, i.e., she will have learned nothing about it.}

\section{Proof}
A key notion in the Mean King problem is the idea of {\em safe vectors} \cite{lit:MeanerKing}. All vectors in the range of a measurement operator $F_x$ must be safe vectors, if Alice does not want to risk a wrong guess. Together with a convenient normalization, we call  $\eta_x$ a safe vector for guessing function $x$ if
\begin{align}\label{gl_bed_safe_vec}
\bracket{\eta_x}{\hat\Phi_b(i)} &= \delta_{x(b),i} \; \forall
\end{align}
with ${\hat\Phi_b(i)}$ from Eq.~(\ref{Phihat}). It is shown in \cite{lit:MeanerKing} that for a non-degenerate choice of bases such a vector exists for every guessing function $x$.

If Alice chooses a measurement $\Menge{F_x}$ that incorporates all of the projectors $p(x)\ketbra{\eta_x}{\eta_x}$, $p(x)\neq 0$ we will call her strategy maximal.
We will now show, that an operator, that possesses all the $\eta_x$ as eigenvectors has to be a multiple of the identity.

\begin{lem}
Let $\hilbertH$ be a Hilbert space of dimension $d$ and $\Menge{\Phi_b(i)}$ a set of  $d+1$ non-degenerate ONBs and let Alice have a maximal successful strategy.
If an operator $E:\hilbertH\otimes\hilbertH\rightarrow\hilbertH\otimes\hilbertH$ fulfills $E\eta_x=e_x\eta_x$ for all safe vectors $\eta_x$ solving (\ref{gl_bed_safe_vec}), then $e_x\equiv e$ is constant and $E=e\hilberteins$.
\end{lem}
\begin{proof}
At first from $\sum_x p(x) \ketbra{\eta_x}{\eta_x} = \hilberteins$ we can conclude that $\spa_\reellR\{\eta_x;\; x\in X\}=\hilbertH\otimes\hilbertH$ holds. Since the number of $d^{d+1}$ guessing functions is larger than the maximal dimension $d^2$ of $\hilbertH\otimes\hilbertH$ and there is a safe vector $\eta_x$ for every guessing function \cite{lit:MeanerKing}, we can ask for a decomposition of a given safe vector $\eta_x$ in a set of linearly independent safe vectors $\Menge{\eta_y}$, with $\eta_x\neq \eta_y$ and  $\eta_x = \sum_y \alpha_y\eta_y$, such that $\alpha_y\neq 0,\; \forall y$. One possibility is to choose the three safe vectors $\eta_u, \eta_v, \eta_w$, whose guessing functions fulfill for two bases $b^\prime,\tilde b\in\intMenge{d+1}$, $b^\prime \neq \tilde b$ the relations
\begin{align*}
x(b^\prime) &= v(b^\prime) = i^\prime\neq j^\prime = u(b^\prime) = w(b^\prime)\\
x(\tilde b) &= u(\tilde b) = \tilde i \neq \tilde j = v(\tilde b) = w(\tilde b)\\
x(b) &= u(b) = v(b) = w(b),\;\; b\notin\{\tilde b, b^\prime\}.
\end{align*}
Then the decomposition is given by $\eta_x = \eta_u + \eta_v - \eta_w$, which can be seen by evaluating the defining equation \ref{gl_bed_safe_vec} for all cases. If $\eta_x,\eta_u,\eta_v,\eta_w$ are eigenvectors of the operator E, it follows from the uniqueness of the decomposition in linearly independent vectors and
\begin{align}
E\eta_x &= E(\eta_u+\eta_v-\eta_w) = e_u\eta_u + e_v\eta_v - e_w\eta_w\\
E\eta_x &= e_x(\eta_u + \eta_v - \eta_w)\label{gl_gleichheitt}
\end{align}
that $\eta_u,\eta_v,\eta_w$ belong to the same eigenvalue.

Now pick two arbitrary safe vectors $\eta_x,\eta_y$ with $x\neq y$, which by the assumptions belong to the eigenvalues $e_x$ and $e_y$. Since $x\neq y$ there exist $1\leqslant m\leqslant d+1$ bases $b_k$ with $x(b)\neq y(b)$.
Now choose a sequence of guessing functions $(z_l)_{l=0}^m$ with $z_0(b)=x(b)$ $\forall b$ and $z_{l+1}(b)=z_l(b)$ $\forall b\neq b_{l+1}$ and $z_{l+1}(b_{l+1})=y(b_{l+1})$ which accounts for $z_m(b)=y(b)$. With the paragraph above it follows, that $e_{z_l}=e_{z_{l+q}}$ for all $l$ and therefore especially $e_x=e$ for all guessing functions $x$ since $x$ and $y$ were arbitrary.

Using that $\spa\Menge{\eta_x}=\hilbertH\otimes\hilbertH$ holds, we can find for all $\Psi\in\hilbertH\otimes\hilbertH$
a set of linearly independent $\Menge{\eta_x}$ such that $\Psi=\sum_x \alpha_x \eta_x$ with $\alpha_x\neq 0$ $\forall x$.
This shows that every $\Psi\in\hilbertH\otimes\hilbertH$ is an eigenvector of E and we can conclude that $E=e\hilberteins$.
\end{proof}

The key to showing security for blocks of length $n$ is to consider these $n$ steps as part of a single instance of a Mean King retrodiction game in a larger Hilbert space with more bases. By tensoring we get $(d+1)^n$ measurement bases $\{\Phi_{\vec b}(\vec i)=\bigotimes_{l=1}^n\Phi_{b_l}(i_l)\}$ on Bob's side in $d^n$ dimensions.
The $n$ entangled states Alice sends to Bob can be considered as an entangled state on $(\hilbertH\otimes\hilbertH)^{\otimes n}$ and apparently a successful strategy is the $n$ times execution of her maximal strategy for the single run $\Menge{F_x}$ obtaining the guessing functions $\vec x = (x_1,\dots ,x_n)$.
The resulting vectors of the form $\eta_{\vec x} = \bigotimes_{l=1}^n \eta_{x_l}$ we will call {\em safe product vectors} since after rearranging of the tensor factors
\begin{align*}
\bracket{\eta_{\vec x}}{\hat\Phi_{\vec b}(\vec i)}=\prod_{l=1}^n \bracket{\eta_{x_l}}{\hat\Phi_{b_l}(i_l)}=\prod_{l=1}^n \delta_{x_l(b_l),i_l}
\end{align*}
they satisfy the constraint \eqref{gl_bed_safe_vec}.

Of course the set of product measurement bases is no longer non-degenerate, but because $\dim (\spa_\reellR(\ketbra{\Phi_{\vec b}(\vec i)}{\Phi_{\vec b}(\vec i)}))= d^{2n}$ holds, they still span the space of all hermitian operators on $\hilbertH^{\otimes n}$ and the same is true for the set of all safe product vectors $\Menge{\eta_{\vec x}}$. From the single safe vectors $\eta_x$, the safe product vectors $\eta_{\vec x}$ inherit the decomposition property, by applying the property to one of the tensor factors.  So every safe product vector $\eta_{\vec x}$ can be decomposed in three linearly independent safe product vectors $\eta_{\vec u}$, $\eta_{\vec v}$ and $\eta_{\vec w}$ with $\eta_{\vec x}=\eta_{\vec u}+\eta_{\vec v}-\eta_{\vec w}$. From this point the argumentation for the operator $E$ follows along the same line as in the single execution case: Firstly for every two different safe product vectors $\eta_{\vec x}$ and $\eta_{\vec y}$ we can find a sequence of guessing functions, that via equation \eqref{gl_gleichheitt} ensure the eigenvalues to be equal. Secondly, since the span of all safe product vectors generates again $\hilbertH\otimes\hilbertH$, which is therefore an eigenspace of the operator $E$ to one single eigenvalue, $E$ is a multiple of the identity. We get

\begin{lem}\label{lem_eigenwert}
Let $\Menge{\eta_{\vec x}}$ be the set of all safe product vectors for the n-times execution of a retrodiction problem with $d+1$ non-degenerate measurement bases $\Menge{\Phi_b(i)}$ on a Hilbert space $\hilbertH$ of dimension $d$, for which Alice has a maximal successful strategy. If $E:(\hilbertH\otimes\hilbertH)^{\otimes n}\rightarrow(\hilbertH\otimes\hilbertH)^{\otimes n}$ fulfills $E\eta_{\vec x}=e_{\vec x}\eta_{\vec x}$ for all $\eta_{\vec x}$ then $e=e_{\vec x}$ for all $x$ and $E=e\hilberteins$ holds.
\end{lem}

We want to analyze the security of the protocol in a transmission-error free scenario for a full coherent attack on both quantum channels. That means, that Eve might eavesdrop on the communication by replacing the transmission line from Alice to Bob by an arbitrary quantum channel $U$  and that her operation $V$ on the feedback channel might be connected to the outcome of this transformation via an additional quantum channel.

\begin{figure}
\centering
\includegraphics[width=0.95\linewidth]{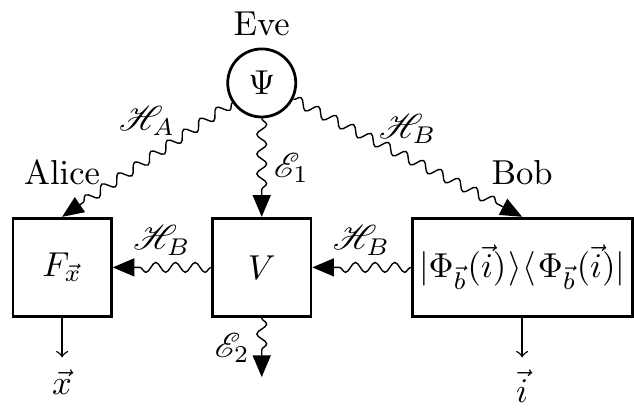}
\caption{Full coherent attack on the protocol}
\label{fig_attack}
\end{figure}

As a further generalization we even give the control of the source of the maximally entangled state to Eve (see FIG. \ref{fig_attack}). In this scenario we proof that if Alice and Bob observe perfect correlations of their data, or, more precisely,  Eve chooses an operation, that does not cause any errors, the resulting key is perfectly secure.

Lemma \ref{lem_eigenwert} shows that the $n$-time execution of a retrodiction game can be viewed as a one time execution of retrodiction game on a larger Hilbert space. Because of this, we can proof the security of the $n$ times execution of the single qudit protocol by proving the security of a protocol where Alice and Bob use a $d^n$ dimensional maximally entangled state initially and the product bases and measurements defined in Lemma \ref{lem_eigenwert}.

Now suppose Eve prepares the state $\Psi\in\hilbertH\otimes\hilbertH\otimes\hilbertE$.
Using $U_{m,l}= X_d^m Z_d^l$ the elements of the generalized Pauli-group of a $d$-dimensional Hilbert space \citep{lit:Bellgen1},
we can decompose this state in the maximally entangled basis on the first two tensor factors and write it as
\begin{align*}
\Psi^{ABE} &=\sum_{m, l, \beta} p_{m, l \beta} (\hilberteins_A\otimes U_{m,l}\otimes\hilberteins_E)\ket{\Omega^{\otimes n}}\otimes\ket{e_\beta}\\
&= \sum_{\beta} (\hilberteins_A\otimes \hat{U}_\beta\otimes\hilberteins_E)\ket{\Omega^{\otimes n}}\otimes\ket{e_\beta}
\end{align*}
where the operators $U_{m,l}$ have the property to generate all basis states of the maximally entangled basis if they are applied in turns to the second tensor factor of $\ket{\Omega^{\otimes n}}$ \citep{lit:Bellgen2}. For the purpose of clarity these $U_{m,l}$ and the $p_{m, l, \beta}$ are absorbed into the $\hat{U}_\beta$. By measuring this state in the product bases $\vec b$ and obtaining the result $\vec i$ Bob projects it onto the state
\begin{align*}
\ket{\Psi_{\vec b, \vec i}}=\sum_{\beta} (\hat{U}_\beta^T\otimes\hilberteins_B\otimes\hilberteins_E)\ket{\hat\Phi_{\vec b}(\vec i)}\otimes\ket{e_\beta}.
\end{align*}
Here we used that $(\hilberteins\otimes \hat{U})\ket{\Omega}=(\hat{U}^T\otimes\hilberteins)\ket{\Omega}$ holds for an Operator $\hat{U}\in\hilbertBOp$ and the maximally entangled state $\ket{\Omega}\in\hilbertH\otimes\hilbertH$.
Eve might implement an arbitrary quantum channel $V$ acting on her subsystem and the eigenstate of the measurement Bob is sending back to Alice. We now analyze the result of this operation:
\begin{align}\label{eq_kanalI}\begin{split}
&V[\ketbra{\Psi_{\vec b, \vec i}}{\Psi_{\vec b, \vec i}}]=\\
&\sum_{\beta,\beta^\prime,l}
(\hat{U}_\beta^T\otimes V_l)\ket{\hat\Phi_{\vec b}(\vec i)}\ket{e_\beta}\bra{\hat\Phi_{\vec b}(\vec i)}\bra{e_{\beta^\prime}}
(\hat{U}_{\beta^\prime}^T\otimes V_l)^\star\otimes\ketbra{e_l}{e_l}
\end{split}
\end{align}
In order to avoid detection, Eve has to restrict her attack to such quantum channels that do not disturb the measurement statistics observed by Alice and Bob. Even after Eve has interfered with the quantum systems Alice should guess the right measurement outcome if no transmission errors are taken into account. This means that the states received by Alice still respond to the same safe vectors $\eta_{\vec x}$ as before so the condition
\begin{align*}
\tr(\ketbra{\eta_{\vec x}}{\eta_{\vec x}}\tr_{E}(V[\ketbra{\Psi_{\vec b, \vec i}}{\Psi_{\vec b, \vec i}}]))=\prod_{l=1}^n \delta_{x_l(b_l),i_l}
\end{align*}
must be respected for all guessing functions $\vec x$, bases $\vec b$ and measurement results $\vec i$.
Expanding the Kraus operators $V_l$ in a standard basis
\begin{align*}
V_l = &\sum_{\gamma,\delta,\mu,\nu} v_{l_{\gamma\delta\nu\mu}} \ketbra{\gamma\delta}{\nu\mu},
\end{align*}
inserting this expression in equation \eqref{eq_kanalI} and tracing out Eve's system we obtain the state controlled by Alice before her final measurement:
\begin{align}\label{gl_rho_Agest}
\rho^{A}_{\vec b, \vec i}=  \sum_{l,k} E_{lk}\ketbra{\hat\Phi_{\vec b}(\vec i)}{\hat\Phi_{\vec b}(\vec i)} E_{lk}^\star
\end{align}
where we defined the operators $E_{lk}$ as
\begin{align*}
E_{lk}=\sum_{\beta} (\hat{U}_\beta^T\otimes(\sum_{\gamma\nu} v_{l_{\gamma k\nu e_\beta}} \ketbra{\gamma}{\nu})).
\end{align*}
Using \eqref{gl_rho_Agest} we find that the probability to measure one of the safe product vectors $\eta_{\vec x}$ for a given tuple of measurement results $\vec i$ in a certain product bases $\vec b$ is given by
\begin{align*}
p(\eta_{ \vec x}, \vec b, \vec i) = \tr(\ketbra{\eta_{\vec x}}{\eta_{\vec x}}\rho^{A}_{\vec b, \vec i})
=  \sum_{l,k} \Betrag{\bracketOp{\eta_{\vec x}}{E_{lk}}{\hat\Phi_{\vec b}(\vec i)}}^2.
\end{align*}
From the constraint $p(\eta_{\vec x}, \vec b, \vec i)=\delta_{\vec x (\vec b), \vec i}$ we can conclude, that $\bracketOp{\eta_{\vec x}}{E_{lk}}{\hat\Phi_{\vec b}(\vec i)}$ has to be valid for every $l$, $k$, $\vec b$, $\vec i$ and $\vec x$. The safe product vectors $\eta_{\vec x}$ are unique one dimensional projectors and $E_{kl}$ cannot have all the $\ket{\hat\Phi_{\vec b}(\vec i)}$ as eigenvectors if $E_{ik}\neq \gamma_{kl} \hilberteins$ holds. This leads to the constraint, that in order to avoid detection, the operators $E_{lk}$ corresponding to Eve's attack have to fulfill the eigenvalue equations
\begin{align*}
E^\star \eta_{\vec x} = \bar \gamma_{lkx} \eta_{\vec x}
\end{align*}
for all $\eta_{\vec x}$. With Lemma \ref{lem_eigenwert} we can conclude, that the operators are of the form $E_{kl}=\gamma_{kl} \hilberteins$ independent of the measurement result $\vec x$ that Alice obtains.

The state under Eve's control after the transmissions is given by:
\begin{align*}
&\rho^E_{final}=\tr_{AB}(V[\ketbra{\Psi_{\vec b, \vec i}}{\Psi_{\vec b, \vec i}}])\\
&= \sum_{l, k, k^\prime} \tr_{AB} (E_{l,k}\ketbra{\hat\Phi_{\vec b}(\vec i)}{\Phi_{\vec b}(\vec i)} E_{l,k^\prime}^\star \otimes\ketbra{k}{k^\prime}\otimes\ketbra{e_l}{e_l}\\
&= \sum_{l, k, k^\prime} \gamma_{k,l}\bar\gamma_{k^\prime,l} \ketbra{k}{k^\prime}\otimes\ketbra{e_l}{e_l}
\end{align*}
Hence $\rho^E_{final}$ is independent of Bob's choice of bases $\vec b$, his measurement result $\vec i$ and Alice guessing function $\vec x$ and Eve cannot infer any information about the exchanged key if she wants to stay undetected.
This shows the security of the proposed protocol in a transmission-error free scenario.

\noindent
\textbf{Acknowledgements: } T.F and A.W. acknowledge funding from the DFG. T.F. is member of Braunschweig IGSM.


\end{document}